\documentclass[amsmath,amssymb,superscriptaddress,aps,a4paper,10pt,prl,twocolumn]{revtex4-1}
\usepackage{dcolumn}

\usepackage[dvips]{color}
\usepackage{graphicx}
\usepackage{amsmath}
\usepackage{amssymb}
\usepackage{amsfonts}
\usepackage{amsthm}
\usepackage{textcomp}
\usepackage{natbib}
\usepackage{gensymb} 
\usepackage[seperr, load={}]{siunitx}
\newcommand{\de}{\mathrm{d}}

\begin{document}


\title{\textbf{Competition between the Superconducting Proximity Effect and Coulomb Interactions in a Graphene Andreev Interferometer}}


\author{Fabio Deon}
\author{Sandra \v{S}opi\'c}
\author{Alberto F. Morpurgo}
	\email{Alberto.Morpurgo@unige.ch}
\affiliation{D\'epartement de Physique de la Mati\`ere condens\'ee (DPMC) and Group of Applied Physics (GAP), University of Geneva, 24 Quai Ernest Ansermet 1211 Gen\`eve 4, Switzerland}


\date{\today}

\begin{abstract}
We have investigated transport through graphene Andreev interferometers exhibiting reentrance of the superconducting proximity effect. We observed a crossover in the Andreev conductance oscillations as a function of gate voltage ($V_{BG}$). At high $V_{BG}$ the energy-dependent oscillation amplitude exhibits a scaling predicted for non-interacting electrons, which breaks down at low $V_{BG}$. The phenomenon is a manifestation of electron-electron interactions, whose main effect is to shorten the single-particle phase coherence time $\tau_\phi$. These results indicate that graphene provides a useful experimental platform to investigate the competition between superconducting proximity effect and interactions.
\end{abstract}


\maketitle


The superconducting proximity effect (PE) consists in the modification of the electronic properties of a normal conductor (N) in contact with a superconductor (S), by induced pair correlations. It is responsible for a broad variety of transport phenomena, such as the the modification of the tunneling density of states in the normal metal \cite{Sueur2008}, the occurrence of Josephson supercurrents \cite{Schapers1997,*Heida1999,*Dubos2001,*DellaRocca2007} and subgap structures \cite{Octavio1983,*Averin1995,*Scheer1997} in superconducting junctions, the enhancement of the subgap conductance in single NS-junctions \cite{Blonder1982}, and the modulation of dissipative transport by the superconducting phase in Andreev interferometers \cite{Pothier1994,*Dimoulas1995,*Petrashov1995}.
Many of these phenomena are by now well understood in terms of Andreev reflection  at the N/S interface, in conjunction with the coherent propagation of non-interacting electron and hole waves in N \cite{Pannetier2000,*Klapwijk2004}, with an overall very good agreement between theory and experiments. On the contrary, except for specific problems (e.g., quantum dots connected to superconductors \cite{DeFranceschi2010}), transport in the regime where electron-electron interactions (EEI) compete with the superconducting PE has remained largely unexplored.

Here we investigate phase-modulated transport through a diffusive Andreev interferometer whose normal region is a T-shaped graphene ribbon. We exploit the possibility to electrostatically tune the transport regime in the ribbon to investigate the competition between the PE and Coulomb interactions. Moving from large carrier density towards charge-neutrality we observe the transition from a regime consistent with expectations based on a non-interacting electron picture, to one in which the PE is increasingly suppressed. On the basis of the experimental observations, we conclude that the suppression is due to EEI, whose main effect appears to be the shortening of the single-particle phase-coherence time $\tau_\phi$ in the N region. Our results indicate that graphene-based hybrid devices provide an excellent platform to explore the PE in the presence of interactions and disorder.

\begin{figure}[t!]
 \centering
 \includegraphics[scale=1,keepaspectratio=true]{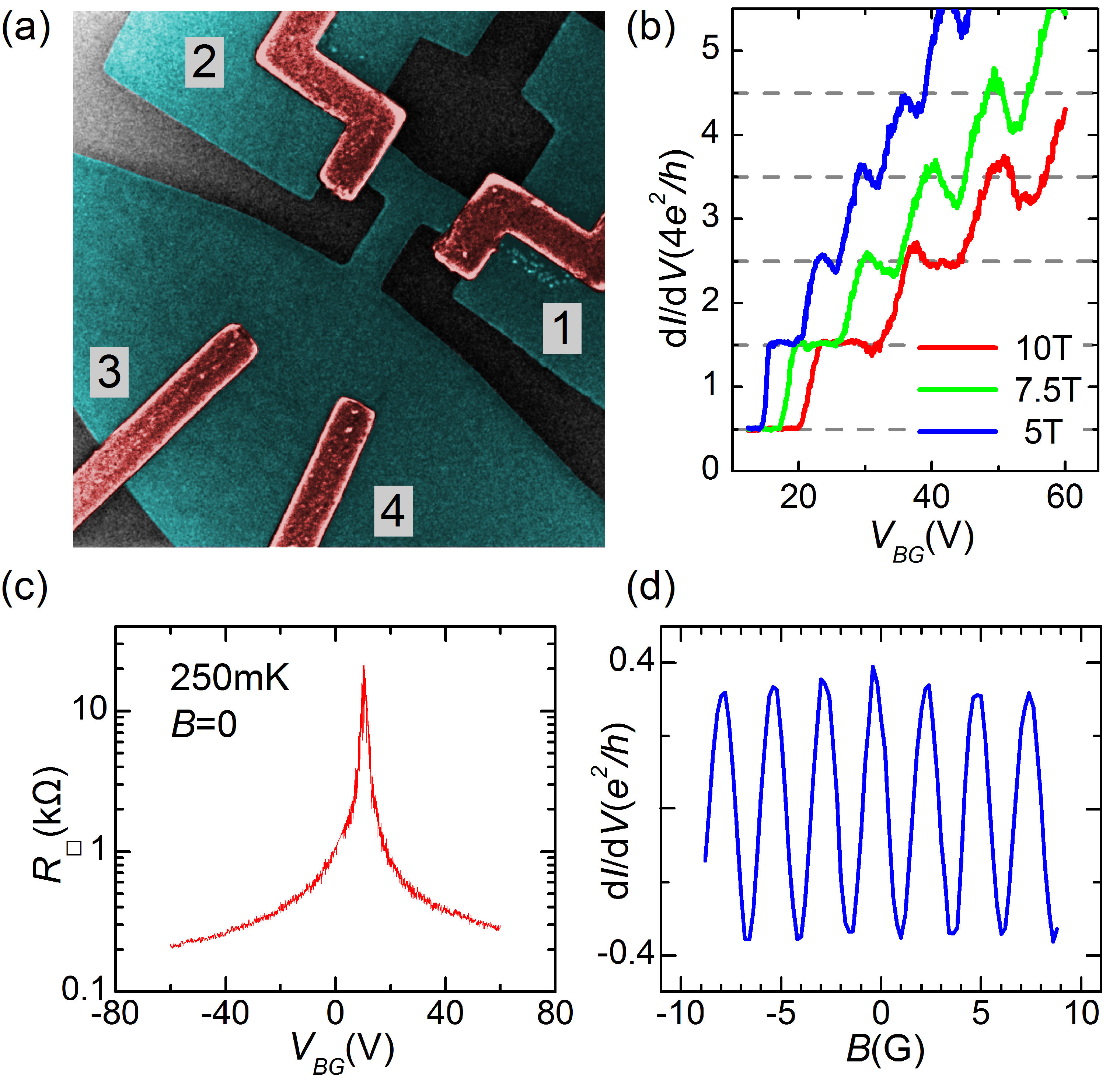}
 \caption{(Color online) (a) False color scanning electron micrograph of one of our  devices; graphene and the superconducting electrodes are colored in cyan and red, respectively (contacts 1 and 2 are joined by a $\simeq \SI{12}{\micro m^2}$ superconducting loop). (b) Two-terminal conductance between contacts 3 and 4 as a function of $V_{BG}$, for different values of perpendicualr magnetic field. (c) Square resistance of the T-shaped graphene ribbon versus $V_{BG}$. (d) Oscillations in the four-probe conductance $G_{3,2|1,4}$, due to quantum interference of Andreev-reflected holes, modulated by the flux threading the loop (data measured at $T=250$ mK with an applied bias $V_{SD}=\SI{40}{\micro V}$) .}
 \label{Fig1}
\end{figure}

Our investigations focus on a specific manifestation of the superconducting PE, namely its so-called reentrance \cite{Artemenko1979,*Beenakker1992,*Lambert1998}. This counter-intuitive phenomenon consists in the non-monotonic energy dependence of the conductance of a diffusive normal metal connected to a superconducting electrode through a highly transparent contact. When the temperature $T$ is lowered from just above the critical temperature $T_C$, the conductance first increases, and then unexpectedly decreases so that at $T=0$ it returns to the normal-state value $G_N$, as if the PE was completely absent. A similar non-monotonic trend is also observed at low temperature, when measuring the differential resistance as a function of applied bias $V$ (i.e., decreasing $V$ from  $V>\Delta/e$ to 0; $\Delta$ is the superconducting gap), and in Andreev interferometers, when looking at the amplitude of the conductance oscillations as a function of bias or temperature.

Theory treating the normal conductor in the diffusive limit relates the energy dependence of the conductance change $\delta G(E)$ to the Thouless energy $E_T=\hbar D/L^2$ (where $L$ is the length of the N region, and $E_T\ll\Delta$) and to the normal state resistance $R_N=1/G_N$  \cite{Nazarov1996,*Stoof1996,*Volkov1996}. For a fixed device geometry, the hallmark of the non-interacting theory is the prediction of a universal scaling of the phenomenon in terms of reduced variables, i.e. when $R_N \delta G(E)$ is plotted as a function of $E/E_T$ \cite{Nazarov1996, *Stoof1996,*Volkov1996}. Pioneering experiments have demonstrated the reentrance of the PE in systems where the normal conductor was either a thin metal film \cite{Charlat1996,*Courtois1999a, Chien1999,*Petrashov1998a} or a two-dimensional electron gas (2DEG) hosted in an InAs or InGaAs-based heterostructure \cite{Hartog1996,*Hartog1996a,*Hartog1997a,Toyoda1999}. Despite quantitative deviations between theory and experiments (especially in the case of InAs-based systems)  attributed to non-ideal aspects of the devices \footnote{In particular, the nature of the contact between the two-dimensional electron gas in InAs and the three-dimensional superconducting contact was thought to be an important cause for quantitative deviations in the amplitude of the measured effects. A similar effect is likely to be at work also in graphene, where at high $V_{BG}$ the device behavior is identical to that observed in InAs-based devices}, the excellent qualitative agreement of the experiments with theoretical predictions has led to the conclusion that the non-interacting theory accounts well for the key aspects of the phenomenon. However, the universality of the scaling between $R_N \delta G(E)$ and $E/E_T$ has never been verified experimentally.

\begin{figure}[t!]
 \centering
 \includegraphics[scale=1,keepaspectratio=true]{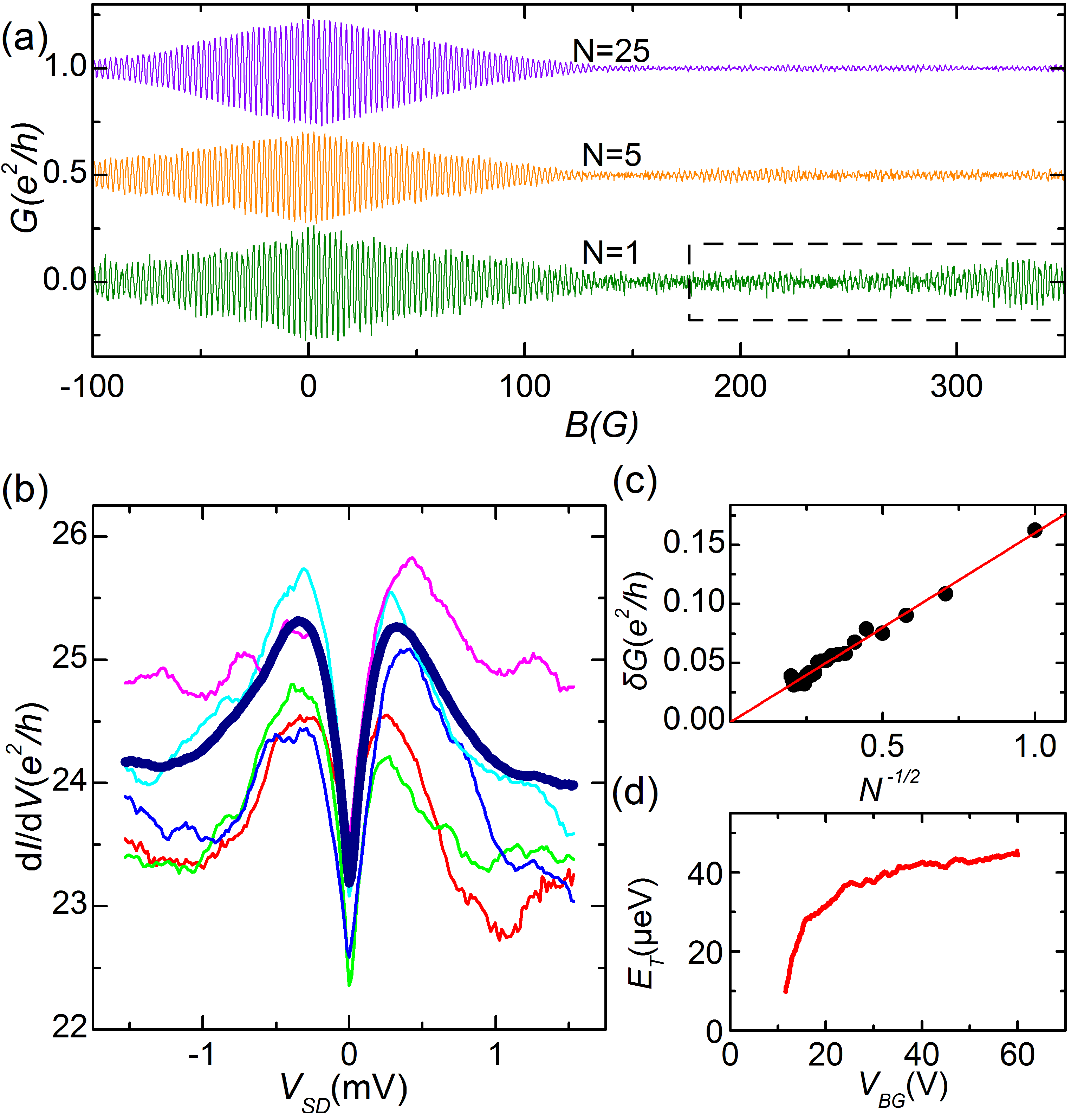}
 \caption{(Color online) (a) progressive averages of the magnetoconductance measured around $V_{BG}=50V$, for $V_{SD}=40\mu V$, as a function of the number $N$ of used to calculate the averages (curves are offset for clarity). (b) Ensemble-averaged (thick line) versus individual $\de I/\de V$ curves (thin lines) measured around $V_{BG}=60V$, at $T=250mK$. (c) Amplitude of the conductance oscillations measured in the range $200G<B<500G$ as a function of the number $N$ of measurements used to average the data. (d) Thouless energy $E_T$ extracted from the $V_{BG}$-dependent resistance of the device.}
 \label{Fig2}
\end{figure}

As compared to devices used in the past, graphene Andreev interferometers offer two main advantages to address the role of interactions. First, the stability and gate tunability of graphene allows us to compare the dependence of the reentrance effect on the transport parameters of the normal conductor. Second, in  graphene ribbons EEI, enhanced by disorder, become relevant as the Fermi energy approaches the charge neutrality point \cite{Han2007,*Liu2009,*Molitor2009,*Han2010,*Oostinga2010,*Gallagher2010}: scattering at the edges increases the tendency of electrons towards  (Anderson) localization, which enhances the effect of Coulomb interactions (indeed, in sufficiently narrow ribbons, fully developed Coulomb blockade is observed \cite{Han2007,*Liu2009,*Molitor2009,*Han2010,*Oostinga2010,*Gallagher2010}). It is this latter point that provides a handle to tune the effect of EEI experimentally.

Figure 1(a) shows a SEM micrograph of one of our Andreev interferometers. A single-layer graphene flake is patterned into a T-shaped ribbon, connected to a superconducting loop, to control the relative phase of the superconducting order parameter $\delta\varphi=2\pi\Phi/\Phi_0$ ($\Phi$ is the magnetic flux threading the loop and $\Phi_0=h/2e$ the superconducting flux quantum). The device is fabricated on graphene exfoliated onto a degenerately doped Si  wafer (coated with a 285 nm thick SiO$_2$ layer) acting as a gate electrode. Two additional probes are placed on the wider graphene region below the vertical arm of the ``T''. The superconducting loop and these electrodes, are defined first (by electron beam lithography, evaporation, and lift-off) and consist of a trilayer of Ti/V/Au (layer thicknesses are 5/17/\SI{5}{nm}; the critical temperature $T_C$ and superconducting gap $\Delta$ are \SI{3.5}{K} and \SI{530}{\micro eV}, respectively). In a second step graphene is etched in an  O$_2$ plasma, through a PMMA mask, to define the T-shaped contour. We have realized and investigated several similar interferometers exhibiting analogous behavior, and here we present data measured on one of these devices.

Experiments were performed in a filtered $^3$He system, down to \SI{250}{mK}. The two-terminal conductance between probes 3 and 4, measured as a function of $V_{BG}$ in the presence of a perpendicular magnetic field $B$ is shown in Fig. 1(b): the observation of clear half-integer quantum Hall plateaus confirms that the device is fabricated on monolayer graphene. The resistance across the ribbon, measured at $B=0$ as a function of $V_{BG}$ (Fig. 1(c)), rises by two-orders of magnitude near the charge neutrality point, due to the mentioned enhanced tendency towards localization. Finally, Fig. 1(d) shows the conductance oscillations induced by a small magnetic field that  modulates the superconducting phase, originating from quantum interference of holes Andreev reflected at the two different superconducting contacts.

As the transport properties of the device are determined by quantum interference, we need to distinguish between contributions of ensemble-averaged (EA) and  sample-specific (SS) nature in the measured quantities. In particular, the SS component of the magnetoconductance oscillations \cite{Hartog1996,*Hartog1996a,*Hartog1997a} can not be neglected at small energies, where the EA component is suppressed by the reentrance of the PE. Therefore, for each quantity of interest we averaged measurements for 25 different values of $V_{BG}$, stepping the gate voltage just enough to cause a change in Fermi energy larger than the correlation energy $\pi^2 E_T$ \cite{Lee1985}. A plot of $E_T=\hbar D/L^2$ for positive $V_{BG}$ is shown in Fig. 2(d). The diffusion constant $D$ is estimated from the zero-bias conductivity, assuming a linear dispersion for graphene \cite{Tombros2007,*Russo2008} (we take $L=\SI{1}{\micro m}$, corresponding to the distance between the bottom of the ``T'' and the superconducting contacts; note how $D$ decreases as the Fermi level approaches charge neutrality, consistently with the expected tendency towards localization).

Figures 2(a,b) show the results of the ensemble averaging performed around $V_{BG}=\SI{60}{V}$, for the magnetic field dependence of the linear conductance and for the bias-dependent differential conductance. Around $B=0$ the amplitude of the conductance oscillations measured at $V_{SD}=\SI{40}{\micro eV}$ is not much affected by the averaging process, because at this bias the EA contribution is larger than the SS one. On the contrary, at higher magnetic field  ($B>\SI{15}{mT}$) where the EA contribution is suppressed by the broken time reversal symmetry (due to the magnetic flux threading the graphene ``T''), averaging over $N$ traces suppresses the amplitude proportionally to $N^{-1/2}$ (see Fig. 2(c)). Similarly, individual $\de I/\de V$ curves are asymmetric and exhibit random bias dependent features, whereas the EA curve is symmetric (thick versus thin lines in Fig. 2(b)). Note how, with decreasing the bias from $V>\Delta/e$, the EA differential conductance $\de I/\de V$  increases in a smooth featureless way, reaches a maximum, and then decreases, so that the value at $V=0$ is comparable to that measured at $V>\Delta/e$. This is a manifestation of the reentrance of the PE.

Having established the averaging procedure, we now look in detail at the EA phase-modulated oscillations at large charge density ($V_{BG}=\SI{60}{V}$). The low-field conductance oscillations are plotted in Fig. 3(a) at $T=\SI{250}{mK}$ and for $V_{SD}$ between $0$ and \SI{0.55}{meV}. Similar measurements have been done as a function of temperature, for $V_{SD}=0$.  The $V_{SD}$- and $T$-dependence of the peak-to-peak amplitude of the first and second harmonic (whose presence is particularly clear at low bias) are shown in Fig. 3(b) and (c). The first harmonic exhibits clear re-entrance both in the bias and in the temperature dependence, with the oscillation amplitude having a maximum at an energy (i.e., either bias or temperature) comparable to the Thouless energy ($E_T\approx\SI{45}{\micro eV}$ at $V_{BG}=\SI{60}{V}$, see Fig. 2(d)). The second harmonic, on the contrary, shows no reentrance. This is expected, because the trajectories causing conductance oscillations with twice the frequency have to Andreev-reflect at both superconducting electrodes, and are therefore longer (by approximately twice the distance between the S contacts). As a result, the effective Thouless energy associated to these trajectories is significantly smaller (by approximately a factor of 5-6) than $E_T$, so that the energy at which reentrance would occur for the second harmonic is smaller than the lowest temperature reached in the experiment \footnote{Virtually all aspects of our measurements far from charge neutrality reproduce what was found in InAs-based interferometers \cite{Hartog1996,*Hartog1996a,*Hartog1997a}, having dimensions, a diffusion constant and carrier density comparable to those of our graphene devices at $V_{BG}=\SI{60}{V}$. This observation directly indicates that the Dirac nature of electrons does not play an important role in disordered graphene on SiO$_2$. On this issue, also see the work by X. Du \emph{et al.} in \cite{Du2008}}.

\begin{figure}[t!]
 \centering
 \includegraphics{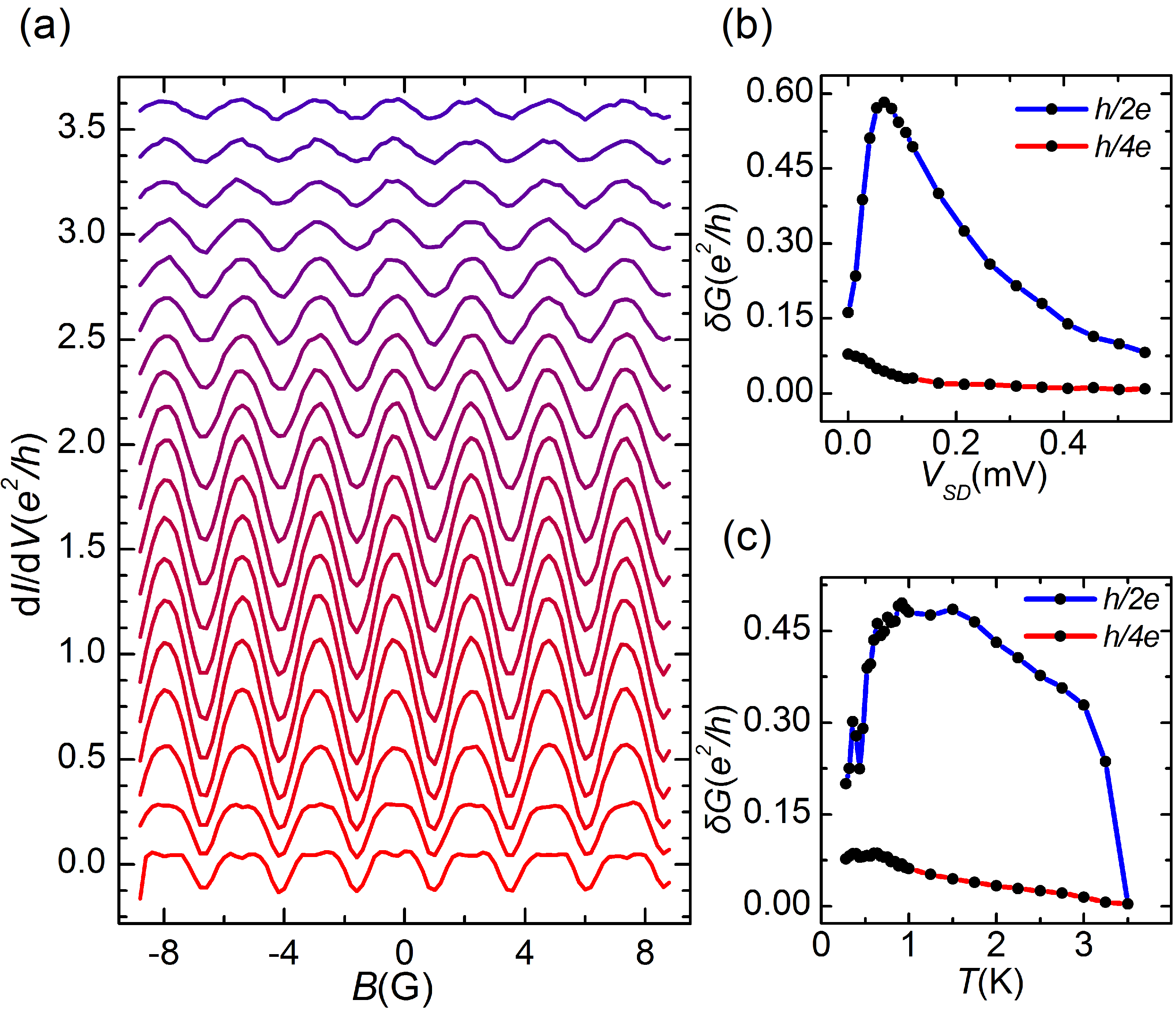}
 \caption{(a) Ensemble-averaged Andreev conductance oscillations measured at $V_{BG}=\SI{60}{V}$ and $T=250$ mK, for $V_{SD}$ varying from 0 (bottom curve) to \SI{0.5}{mV} (top; curves offset for clarity). (b),(c) Bias and temperature dependence of the amplitude of the first and second harmonics of the oscillations shown in (a).}
 \label{Fig33}
\end{figure}

We now analyze the evolution of the energy dependence of the oscillations as a function of $V_{BG}$. Fig. 4(a) shows the bias dependence of the EA oscillation amplitude (first harmonic) for seven different values of  $V_{BG}$ between \SI{60}{V} and \SI{12.5}{V}. Upon lowering  $V_{BG}$, the maximum oscillation amplitude decreases, qualitatively in line with expectations based on a non-interacting theory, because $R_N$ increases (see Fig. 1(c)). The value of $V_{SD}$ corresponding to the maximum oscillation amplitude, however, remains unexpectedly unchanged. Within a non-interacting picture, this finding is inconsistent with the value of the Thouless energy $E_T$, which changes from $\sim\SI{45}{\micro eV}$ to $\sim\SI{10}{\micro eV}$ as $V_{BG}$ is lowered (Fig. 2(d)). For a more quantitative analysis, we look at the data in terms of normalized quantities, i.e. we plot $R_N \delta G$ as a function of $eV_{SD}/E_{T}$ (see Fig. 4(b)). At large densities, specifically for $V_{BG}=60$, 50 and \SI{40}{V}, the rescaled curves fall on top of each other, as expected in the absence of EEI. Starting from $V_{BG}=\SI{30}{V}$, however, deviations from perfect scaling become progressively larger as $V_{BG}$ is lowered. Specifically, the maximum relative oscillation amplitude decreases, and shifts to larger $eV_{SD}/E_{T}$ ratios. Both trends observed in the range $\SI{12.5}{V}<V_{BG}<\SI{30}{V}$ are in conflict with what is expected for a non-interacting system, and indicate the occurrence of a crossover to a different transport regime.

\begin{figure}[t!]
 \centering
 \includegraphics{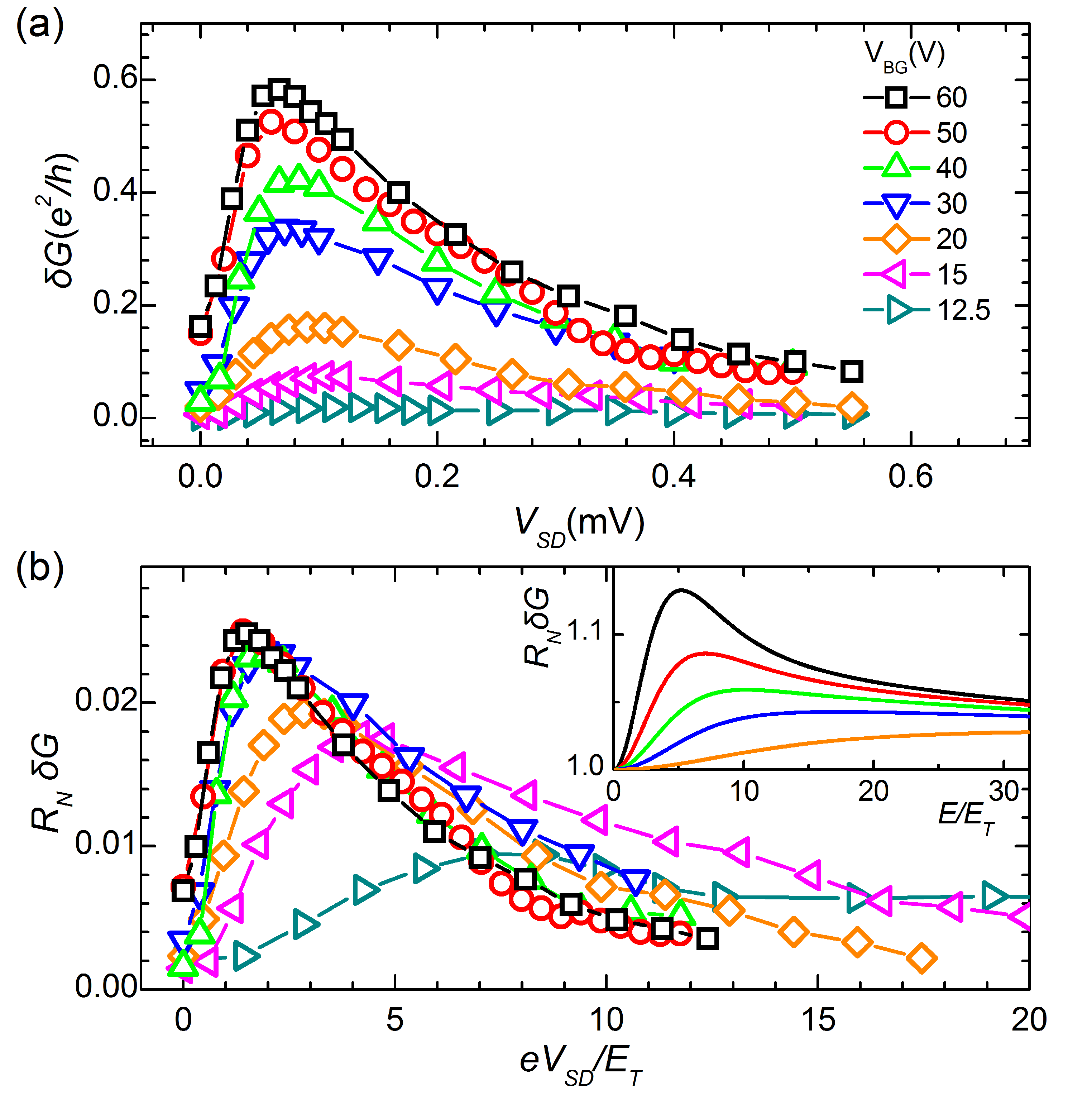}
 \caption{(a) Bias dependence of the amplitude of  the ensemble averaged Andreev conductance oscillations (first harmonic) measured at $T=\SI{250}{mK}$ near zero magnetic field ($\SI{-10}{G}<B<\SI{10}{G}$), for seven different values of $V_{BG}$ between 60 and \SI{12.5}{V} (as indicated in the legend; charge neutrality is at $V_{BG}=\SI{10.5}{V}$). (b) Same data as in (a), plotted in  dimensionless units: the curves measured for $V_{BG}= 60, 50,$ and \SI{40}{V} exhibit a perfect scaling, which breaks down starting from $V_{BG}=\SI{30}{V}$. Inset: calculated conductance of a diffusive NS junction for increasingly shorter values of $L_\phi/L$ ($\infty$, black curve, 0.6, 0.4, 0.3, 0.2, orange curve), using the linearized Usadel equations, as in \cite{Charlat1996,*Courtois1999a}. Note how the maximum decreases in amplitude and shifts to larger energies with decreasing $L_\phi$.}
 \label{Fig4}
\end{figure}

The evolution of the bias-dependent differential conductance $\de I/\de V(V_{SD})$ with  $V_{BG}$ (see Fig. 5(a-c)) provides additional information. At $V_{BG}=60$ V (top panel) Andreev reflection results in a clear conductance increase. At higher temperature (\SI{3.5}{K}) the conductance enhancement extends to all subgap voltages while at low $T$ a conductance dip appears at low bias (i.e. the phenomenology of the reentrance effect). Upon lowering $V_{BG}$ to \SI{20}{V} (Fig. 5(b)), the conductance enhancement at subgap voltage becomes significantly less pronounced; eventually, for $V_{BG}$ sufficiently close to charge neutrality ($V_{BG}=\SI{12.5}{V}$, Fig. 5(c)) no enhancement of $\de I/\de V(V_{SD})$ is observed, and only a suppression persists, which occurs on an energy scale larger than the superconducting gap. This suppression is what is typically seen in low-dimensional systems where dynamical Coulomb blockade becomes relevant \cite{Egger2001,*Bachtold2001,*Kanda2004,*Nazarov2009}, and is expected in wide graphene ribbons at temperatures for which a full transport gap due to disorder-induced Coulomb blockade is not fully developed. We conclude that the manifestation of EEI in the $\de I/\de V(V_{SD})$ curves occurs over the same $V_{BG}$ range for which the maximum amplitude of the conductance oscillations decreases rapidly and shifts to higher $E/E_T$ ratios (see Fig. 4(b)). For these reasons, we attribute the observed deviations in the scaling between $R_N \delta G(E)$ and $E/E_T$ to EEI.

\begin{figure}[b!]
 \centering
 \includegraphics{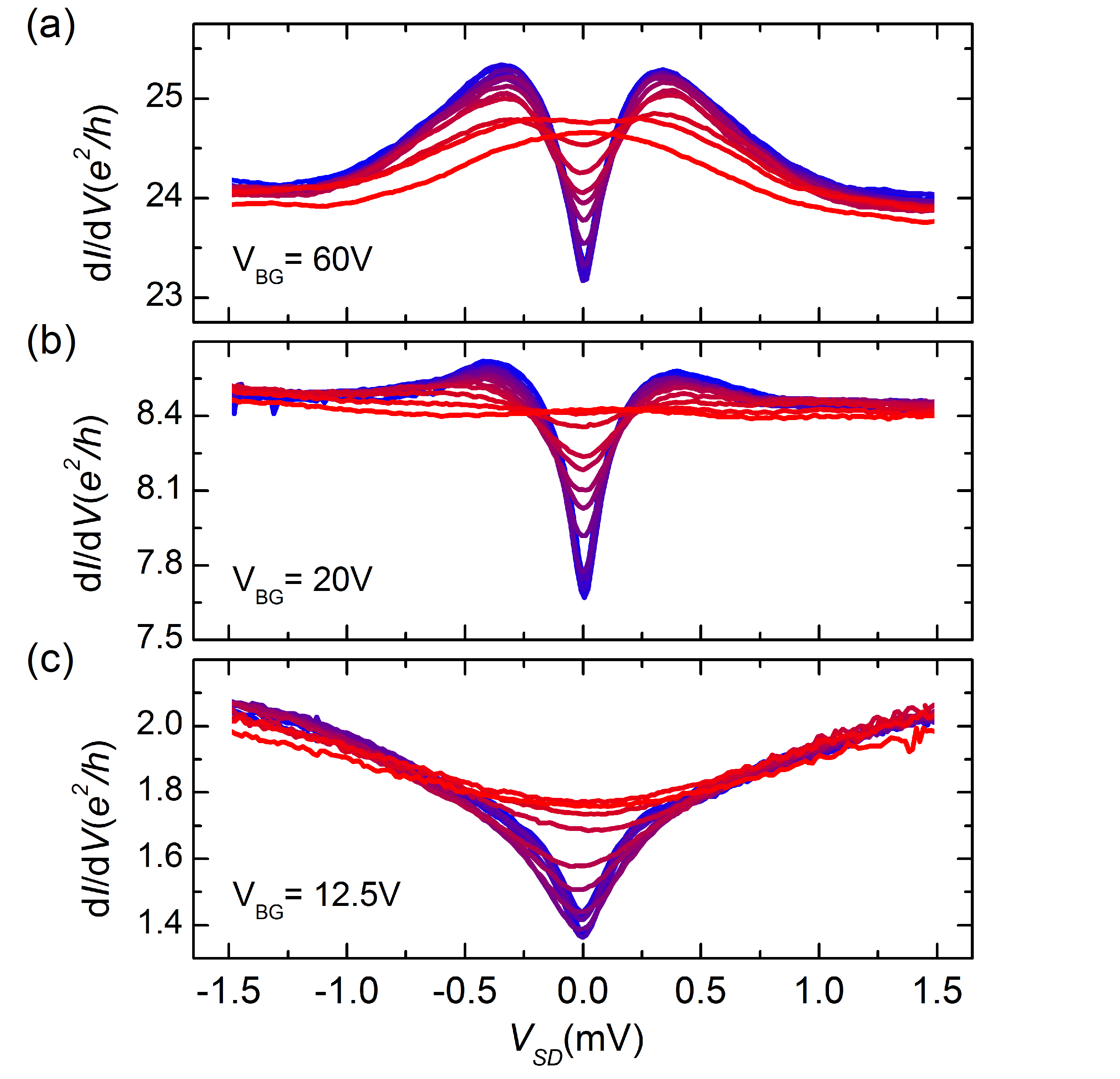}
 \caption{(a)-(c) Ensemble-averaged differential conductance curves between \SI{265}{mK} (blue) and \SI{3.5}{K} (red), for different values of $V_{BG}$ (60,20, and 12,5 V, respectively). With approaching charge neutrality, the conductance enhancement due to Andreev reflection visible in (a) is suppressed (b), and eventually completely disappears (c).}
 \label{Fig5}
\end{figure}

It is known theoretically that Coulomb interactions tend to suppress pairing correlations and compete with the superconducting proximity effect. In a context very closely related to the one discussed here -- the so-called reflectionless tunnelling \cite{Kastalsky1991,*Wees1992}, i.e. the manifestation of the proximity effect in devices analogous to ours, but with the N and S regions that are tunnel coupled -- the suppression of proximity effect by EEI has been analyzed theoretically \cite{Huck1998,*Oreg1999,*Semenov2012}. It was found that the influence of EEI on the electronic phase coherence plays an important role. Indeed, the damping and shift towards higher energy of the maximum in the Andreev oscillations observed in our experiments can be accounted for in terms of a progressive shortening of the single-particle coherence time $\tau_{\phi}$. Specifically, a finite value of the phase-beaking length $L_\phi=\sqrt{D\tau_\phi}$ introduces a cutoff for the penetration of the pair amplitude in N: when $L_\phi$ drops  below $L$, $E_\phi=\hbar D/L_\phi^2$ takes the role of $E_T$ in determining the energy-scale of the reentrance. Having a new energy scale (next to $E_T$) that becomes relevant explains the deviations from scaling on the energy axis. Dephasing obviously also explains why the amplitude of the proximity effect decreases, since trajectories whose length is larger than $L_{\phi}$ cannot contribute to phase coherent effects. The effect of shortening $\tau_{\phi}$ is illustrated in the inset of Fig. 4(b). The plots show the conductance of a single NS junction, calculated by solving the linearized Usadel equations \cite{Charlat1996,*Courtois1999a} for five different values of the phase coherence time: the trends observed reproduce qualitatively the behavior of the Andreev conductance oscillations. At the conceptual level our observations appear to be in line and to support the conclusions of theoretical studies \cite{Huck1998,*Oreg1999,*Semenov2012}.

In conclusion, we have used graphene-based Andreev interferometers to investigate the competition between the superconducting PE and electron-electron interactions, which manifests itself through a crossover in the behavior of the Andreev conductance oscillations as a function of gate voltage. Interaction effects become more pronounced as the Fermi level is shifted closer to the charge neutrality point, and can be interpreted in terms of a progressive suppression of phase coherence in graphene. 
Earlier experiments had already shown that graphene is a versatile experimental platform for the investigation of proximity-induced superconductivity in conductors with tunable transport properties \cite{Heersche2007,*Du2008,*Dirks2011,*Allain2012}. Our results now demonstrate that hybrid devices based on graphene nanoribbons are also particularly suitable to study the interplay between electron-electron interactions and the superconducting proximity effect.

\begin{acknowledgments}
We thank A. Ferreira for technical support. Financial support from the ESF ENTS project, SNF and NCCR-QSIT is gratefully acknowledged.
\end{acknowledgments}



%

\end{document}